\documentclass[11pt,aps,prb,floatfix,nofootinbib]{revtex4}

\usepackage{amsmath}
\usepackage{graphicx}
\usepackage{color}

\usepackage{dingbat}
\usepackage{amssymb}
\usepackage{float}

\usepackage{fancyhdr}
\usepackage{graphicx}
\usepackage{amssymb} 
\usepackage{url}     
\usepackage{bm}      

\makeatletter
	\renewcommand\@biblabel[1]{#1.}
    \def\@cite#1#2{$^{\mbox{\scriptsize #1\if@tempswa , #2\fi}}$}
\makeatother

\begin{document}
\renewcommand*{\thefootnote}{\fnsymbol{footnote}}

\thispagestyle{empty}
\begin{center}
{\Large {\bf Optoelectronics with electrically tunable PN diodes in a monolayer dichalcogenide}}\\
\quad \\
{\large Britton W.~H.~Baugher\footnote{These authors contributed equally to this work.}, Hugh O.~H.~Churchill$^*$, Yafang Yang, and Pablo Jarillo-Herrero} \\
{\it Department of Physics, Massachusetts Institute of Technology, Cambridge, MA 02139, USA} \\
\end{center}
\date{\today}

\textsf{\textbf{
One of the most fundamental devices for electronics and optoelectronics is the PN junction, which provides the functional element of diodes, bipolar transistors, photodetectors, LEDs, and solar cells, among many other devices.
In conventional PN junctions, the adjacent p- and n-type regions of a semiconductor are formed by chemical doping.  Materials with ambipolar conductance, however, allow for PN junctions to be configured and modified by electrostatic gating.
This electrical control enables a single device to have multiple functionalities.
Here we report ambipolar monolayer WSe$_2$ devices in which two local gates are used to define a PN junction exclusively within the sheet of WSe$_2$.  With these electrically tunable PN junctions, we demonstrate both PN and NP diodes with ideality factors better than 2.  Under excitation with light, the diodes show photodetection responsivity of 210 mA/W and photovoltaic power generation with a peak external quantum efficiency of 0.2\%, promising numbers for a nearly transparent monolayer sheet in a lateral device geometry.  Finally, we demonstrate a light-emitting diode based on monolayer WSe$_2$.  These devices provide a fundamental building block for ubiquitous,  ultra-thin, flexible, and nearly transparent optoelectronic and electronic applications based on ambipolar dichalcogenide materials.
}}

Next generation photodetectors and photovoltaic devices, as well as sensors, displays, and light emitting diodes, will all require new optoelectronic materials with superior characteristics to those currently in use.  Candidate materials must be flexible for wearable devices, transparent for interactive displays, efficient for solar cells, and robust and low cost for broad distribution.  Monolayer semiconducting transition metal dichalcogenides (TMDs) such as tungsten diselenide (WSe$_2$) are flexible\cite{Bertolazzi:2011}, nearly transparent\cite{Mak:2010,Chuang:2013}, high strength\cite{Bertolazzi:2011}, direct band gap\cite{Mak:2010} materials that have the potential to meet all of these criteria.
Tungsten diselenide's crystal structure is comprised of stacks of trilayer sheets made of a single atomic layer of tungsten encapsulated by two layers of selenium.  This structure leads to very strong intralayer bonding in the trilayer plane and weak interlayer bonding out of plane.  Such bonding allows for the exfoliation of bulk crystals down to a single molecular layer of Se-W-Se in a manner similar to graphene\cite{Novoselov:2005}.

Monolayer WSe$_2$ is a direct gap semiconductor with a band gap of $\sim 1.65$ eV (ref. \citenum{Zhao:2013}), a complement to other two-dimensional materials\cite{Geim:2013} such as graphene, a gapless semimetal, and boron nitride, an insulator.  The direct band gap distinguishes monolayer WSe$_2$ from its bulk and bilayer counterparts, which are both indirect gap materials with smaller band gaps\cite{Wilson:1969, Zhao:2013, Ding:2011}.  This sizable direct band gap in a two-dimensional layered material enables a host of new optical and electronic devices for both applications and fundamental investigation.  Thin film TMDs have already demonstrated novel nanoelectronic and optoelectronic devices such as ambipolar and high quality field effect transistors\cite{Podzorov:2004,Radisavljevic:2013,Baugher:2013}, electric double-layer transistors\cite{Zhang:2013}, integrated circuits\cite{Wang:2012}, and phototransistors\cite{Yin:2012} with high responsivity\cite{LopezSanchez:2013}.  In addition, photoluminescence\cite{Mak:2010, Splendiani:2010} and electroluminescence\cite{Sundaram:2012,Ye:2013} have been used to observe effects such as giant spin-valley coupling\cite{Zeng:2013} and optical control of valley polarization \cite{Mak:2012, Sallen:2012} and coherence \cite{Jones:2013}.

PN junctions in TMDs have, to date, been much less explored than field effect transistors.
Junctions of p- and n-doped bulk WSe$_2$ were reported over 30 years ago \cite{Spah:1983}.  But only in the last few years have thin film TMDs been made into various types of PN devices, such as an ionic-liquid gated bulk MoS$_2$ device\cite{Zhang:2013}, an InAs - bulk WSe$_2$ heterojunction \cite{Chuang:2013}, and a doped silicon - monolayer MoS$_2$ heterojunction \cite{Ye:2013}.
Unfortunately, bulk TMDs and heterostructures involving other materials lack many of the most appealing properties of monolayer TMDs:  a direct band gap, flexibility, and transparency.
Here we present measurements of monolayer WSe$_2$ PN junctions defined and controlled by electrostatic gates.  We demonstrate PN junction devices with diverse functionality, including current rectifying diodes, a photodetector, a photovoltaic device, and a light-emitting diode, and we characterize their performance.

Device fabrication begins with the exfoliation of natural crystals of WSe$_2$ onto glass/PDMS/MMA transfer slides.  Monolayer flakes are identified by optical contrast\cite{Benameur:2011} and then aligned and transferred\cite{Dean:2010,Zomer:2011} onto substrates with a pre-patterned pair of local back gates separated by 100 nm and covered by 20 nm of HfO$_2$.  The flakes are then contacted by gold.  An optical image and schematic of the device used for all measurements except electroluminescence is shown in Fig.~1a.  More comprehensive fabrication details are provided in the Methods and Supplementary Information.
All measurements were performed in vacuum ($\sim 10^{-5}$ Torr) and at room temperature.
For all transport measurements we report, a voltage bias, $V_{ds}$, is applied to the source contact and the DC current through the device, $I_{ds}$, is measured at the drain.

\begin{figure}[h]
\centering
\includegraphics[width=5.25in]{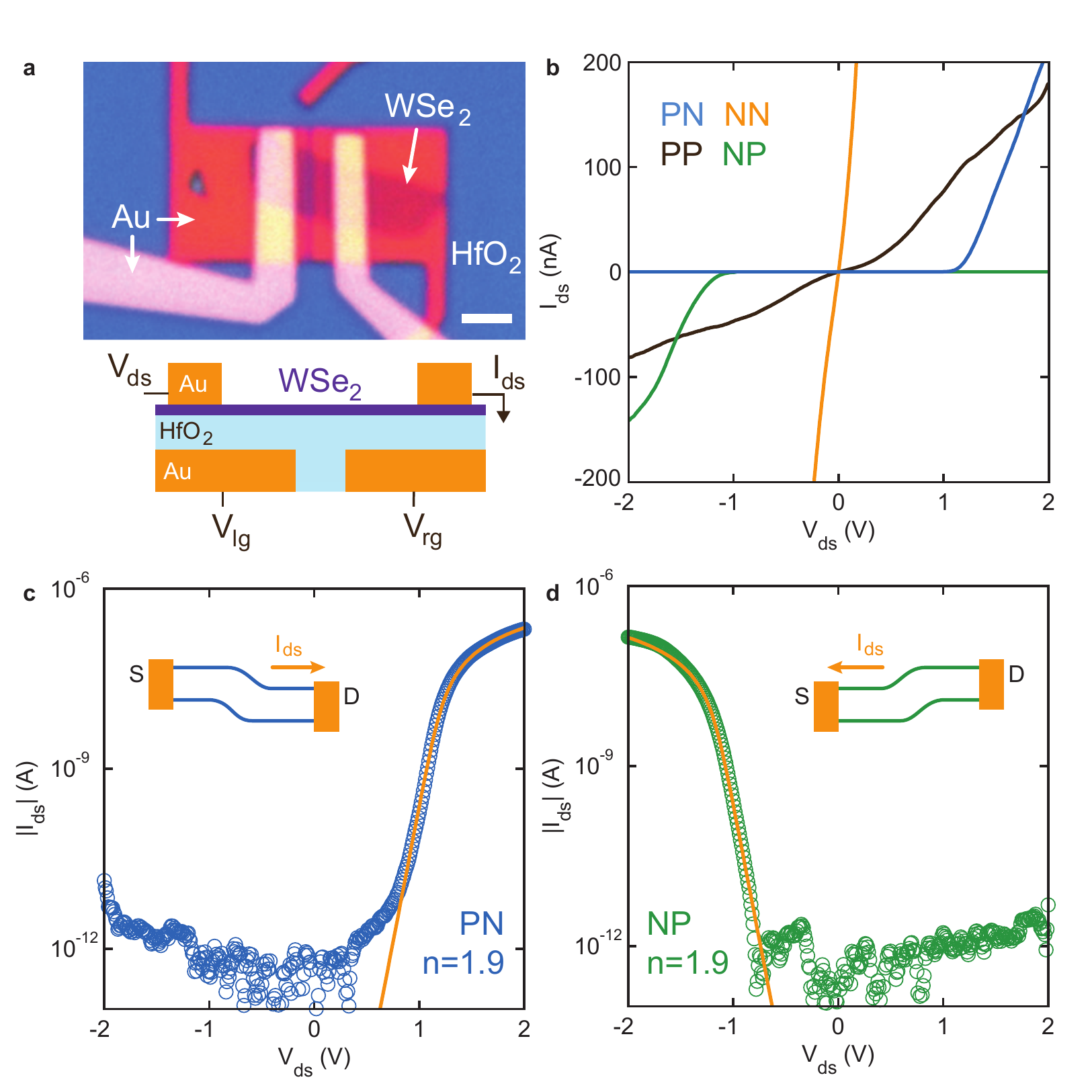}
\caption{
\textsf{\textbf{Gate-controlled monolayer WSe$_2$ PN junction diodes.  a,} Optical micrograph of a monolayer WSe$_2$ device controlled by two local gates.  The WSe$_2$ is contacted with Au electrodes.  The flake and contacts are insulated from the gates by 20 nm of HfO$_2$.  The scale bar is 2 $\mu$m.  Below is a schematic side view of the device including electrical connections.  \textbf{b,}  Current-voltage ($I_{ds}$-$V_{ds}$) curves showing transport characteristics of the four doping configurations of the device, NN, PP, PN, and NP.  Both gates were set to 10 V for the NN configuration and -10 V for PP.  $V_{lg}$ was set to $\pm 10$ V and $V_{rg}$ to $\mp 10$ V for PN/NP.  The NN and PP configurations (yellow and black curves respectively) are Ohmic at low $V_{ds}$, while the PN and NP diodes (blue and green curves, respectively) strongly rectify current in opposite directions.  \textbf{c, d} Semi-log plots of $I_{ds}$ through the PN (blue circles) and NP (green circles) diodes as a function of $V_{ds}$ with fits in yellow (see text).  The fits give a diode ideality of $n=1.9$ for both the PN and NP configurations.  Insets: Schematic band diagram of the device in forward bias for PN and NP configurations.
}}
\end{figure}

The voltages on the two gates, $V_{lg}$ for the left gate and $V_{rg}$ for the right gate, independently control the carrier density in the left and right sides of the monolayer.
In this way the device can be electrostatically doped into various conducting regimes (Fig.~1).
With both sides of the device n-doped ($V_{lg} = V_{rg} = 10$ V, denoted NN) or both sides p-doped ($V_{lg} = V_{rg} = -10$ V, denoted PP), the device shows an Ohmic current-voltage relation at low $V_{ds}$ (Fig. 1b).  The device has a significantly higher conductance in the NN configuration, most likely due to lower contact resistance between the gold and the WSe$_2$ for n-type conductors, as has been observed in MoS$_2$ (Ref. \citenum{Radisavljevic:2011}).

By oppositely biasing the two gates, doping one side with electrons and the other with holes, the device rectifies the flow of current as a diode.
Current measurements for the two diode configurations ($V_{lg} = -10$ V, $V_{rg} = 10$ V, denoted PN and $V_{lg} = 10$ V, $V_{rg} = -10$ V, denoted NP) are shown in Fig. 1b.
By selecting the PN or NP configuration, the device can rectify current flowing in either direction.
The drain-source currents of the diodes are fit (Figs.~1c and 1d) to an extension\cite{Banwell:2000} of the Shockley diode equation \cite{Sah:1957} that includes a series resistance, $R_s$:
$$I_{ds}=\frac{nV_T}{R_s}W\left[ \frac{I_0R_s}{nV_T}\exp\left(\frac{V+I_0R_s}{nV_T}\right)\right]-I_0$$
where $V_T=k_BT/q$ is the thermal voltage at a temperature $T$, $k_B$ is the Boltzmann constant, and $q$ is the electron charge.  $I_0$ is the reverse bias current, and $n$ is the diode ideality factor, where $n=1$ denotes an ideal diode.  The $W$ function is the Lambert $W$ function\cite{Lambert:1758}.

The fits give ideality factors of $n = 1.9$ for both the PN and NP configurations, indicating that the current is mostly limited by recombination rather than diffusion \cite{Sah:1957}.
The series resistances extracted from the fits are $R_s = 3 M\Omega$ for PN and $R_s = 5 M\Omega$ for NP.
We note that our data do not strongly constrain the value of the reverse-bias saturation current because it is below the 1 pA noise floor of our measurements.  An increase in current at high reverse bias indicates a shunt resistance across the junction of order 0.5 T$\Omega$.
Both diodes have rectification factors of $10^5$ and reverse bias currents less than 1 pA up to $|V_{ds}|=1$ V, which are promising characteristics for low power electronics.

A more complete view of transport through the device is shown in a map of current as a function of the two independent back gates (Fig.~2).  The four corners of the map show the extremes of the four doping configurations available with two gates (NN, PP, PN, and NP). The off state of the device due to its band gap can be seen in the dark blue region in the center of the map, separating the four conducting configurations.
Though mid-gap states, most likely due to disorder, form a conducting region between the NN and NP quadrants, current through these states is thermally activated and can be eliminated by cooling the device to 200 K (see Supplementary Information).
With both gates grounded, the device is an insulator, indicating a low level of extrinsic doping of this natural WSe$_2$, in contrast with the high level of n-doping frequently observed in natural MoS$_2$ (ref. \citenum{Novoselov:2005}).  This allows for ambipolar conduction through the device.  Along the diagonal line defined by $V_{bg}=V_{lg}=V_{rg}$, line cuts show the device behaving as an ambipolar field effect transistor (Fig.~2c).

\begin{figure}[h]
\centering
\includegraphics[width=5.25in]{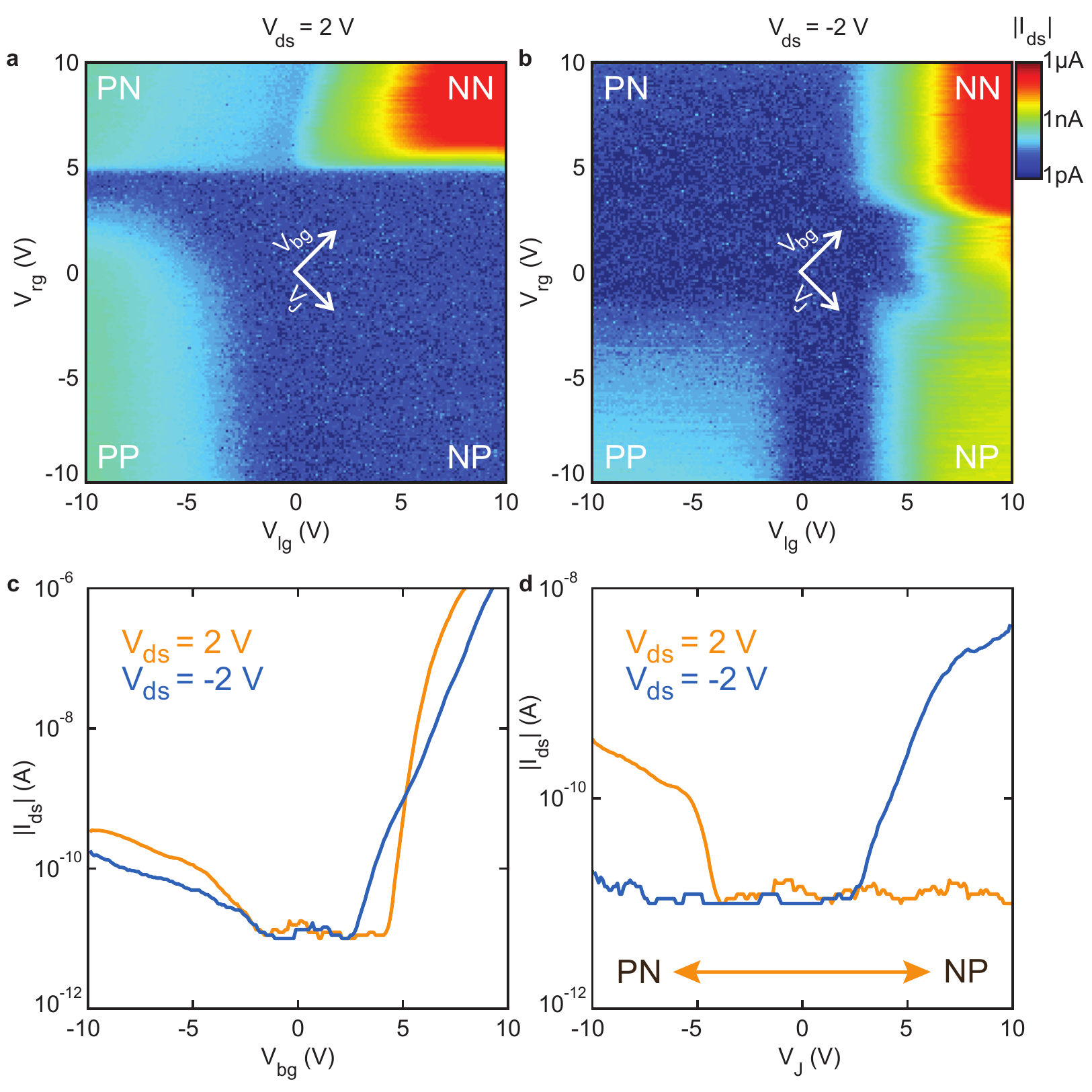}
\caption{
\textsf{\textbf{Current through the device as a function of doping configuration.  a, b,}  Current magnitude through the device, $|I_{ds}|$, as the drain-source bias, $V_{ds}$, is held at $2$ V ({\bf a}) and $-2$ V ({\bf b}) and the two back gates are varied independently.  The color scale is logarithmic from 1 pA to 1 $\mu$A.  \textbf{c,} Diagonal cuts of $|I_{ds}|$ where both gates are swept together as one back gate, $V_{bg} = V_{lg} = V_{rg}$, for $V_{ds}=2$ V (yellow curve) and $-2$ V (blue curve).  These curves are characteristic of an ambipolar field-effect transistor.  \textbf{d,} Off-diagonal cuts of $|I_{ds}|$ where the two gates are swept with opposite polarity, showing the dependence on the asymmetric gate voltage, $V_J = V_{lg}=-V_{rg}$, that defines the junction.  The current was measured at $V_{ds}= 2$ V (yellow curve) and $-2$ V (blue curve), demonstrating rectification of current in opposite directions for the PN ($V_J < 0$) and NP ($V_J > 0$) gate configurations.
}}
\end{figure}

With the gate voltages adjusted to have opposite signs, the current maps in Figs.~2a and 2b show the current in the PN and NP configurations.  For positive $V_{ds}$, current flows freely in the forward-biased PN region (upper-left corner) and is suppressed in the reverse-biased NP region (lower-right corner).  When the bias is reversed, this rectifying behavior is also reversed.  Along the off-diagonal, line cuts show gate-controlled rectifying behavior (Fig. 2d) as a function of the asymmetric gate voltage, $V_J = V_{lg}=-V_{rg}$, that defines a junction between the p- and n-type regions.  We note here that there was a decrease in device performance between the datasets presented in Fig.~1 and Fig.~2, resulting in lower conductance in Fig.~2.

To investigate the optoelectronic properties of monolayer WSe$_2$ PN diodes we begin by using scanning photocurrent microscopy (Fig.~3a) to measure current through the device as a laser spot ($\approx 2 \mu$m diameter) is scanned over the sample.
With the gates set to the NP configuration, photons impinging on the junction create electron-hole pairs that are separated to opposite contacts by the electric field at the junction, generating a photocurrent. By scanning the beam (75 $\mu$W, 830 nm diode laser) over the sample and measuring drain current at $V_{ds}=0$, we obtain a spatial map of the photoresponse of the junction (Fig. 3b).  Calibrating the photocurrent map with a simultaneously acquired image of reflected light from the sample (see Supplementary Information), we overlay the positions of the contacts and gates to demonstrate that photocurrent arises when the light hits the junction.  A line cut through the center of the photocurrent map is shown in Fig.~2c with the positions of the contacts and the junction illustrated for reference.

\begin{figure}[h]
\centering
\includegraphics[width=5.25in]{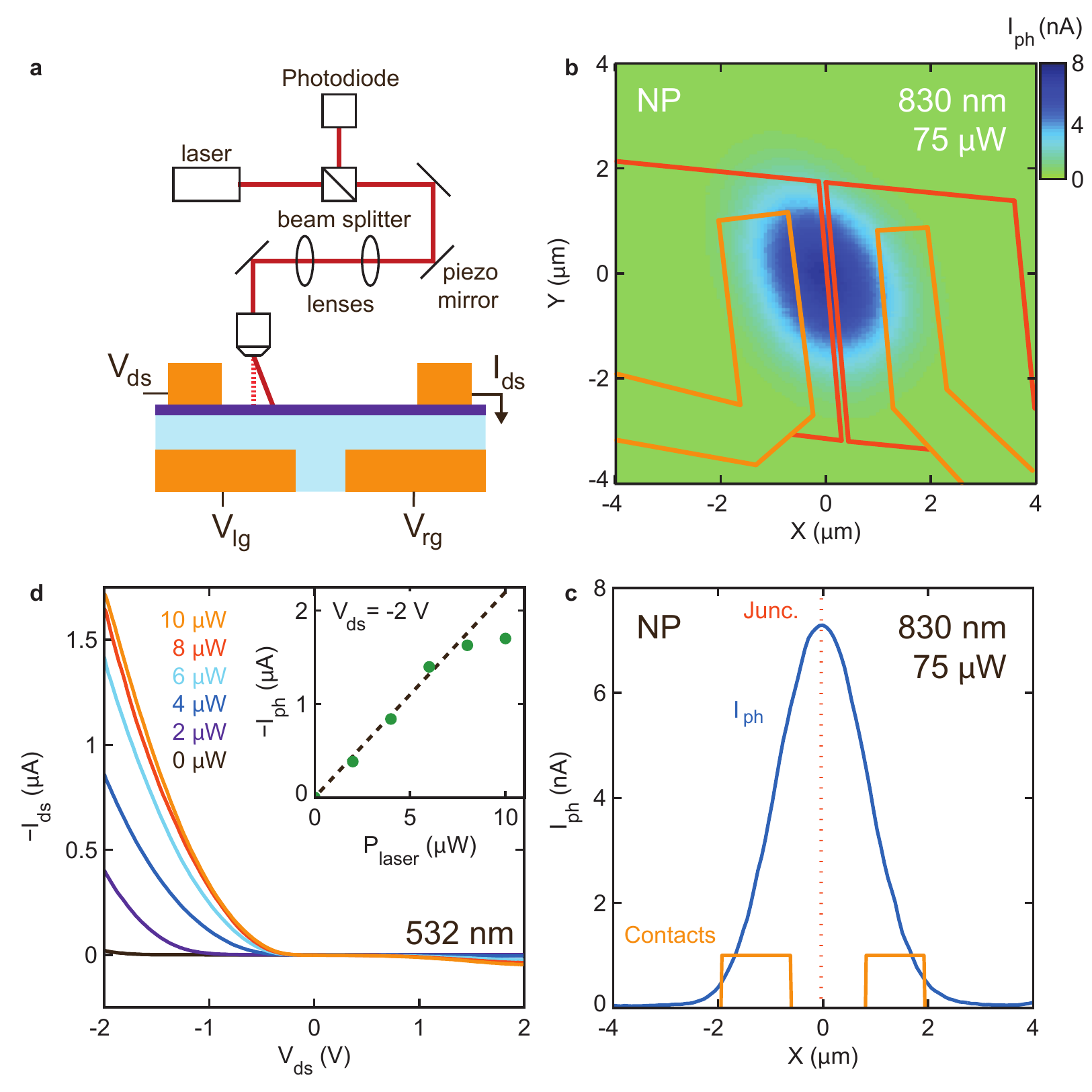}
\caption{
\textsf{\textbf{Photodetection in monolayer WSe$_2$.  a,} Schematic of the scanning photocurrent microscopy setup and the device.  The laser is focused onto the sample through a microscope objective at a position set by a piezo-controlled mirror.  As the laser is scanned over the sample, photocurrent is recorded simultaneously with the reflected laser power measured by a photodiode.  \textbf{b,} Photocurrent, $I_{ph}$, as a function of laser position (75 $\mu$W, 830 nm diode laser), with the device in the NP configuration and $V_{ds}=0$ V.  The peak in current corresponds to the location of the junction.  Outlines of the contacts and gates are overlaid on the current map based on the reflected image (see Supplementary Information).  \textbf{c,} Line cut from {\bf b} at Y=0.  Outlines of the contacts and the position of the junction are show in yellow and orange based on the reflected image.  \textbf{d,} $I_{ds}$-$V_{ds}$ characteristics with the device in the NP configuration at laser powers $0-10 \mu$W from a 532 nm diode laser.  Inset: $I_{ph}$ (green dots) at $V_{ds} = -2$ V as a function of laser power.  A linear fit (black dashed line) to $I_{ph}$ for powers $0-8 \mu$W gives a responsivity of $210$~mA/W.
}}
\end{figure}

At relatively large $V_{ds}$ while in the NP gate configuration, a substantial photocurrent is generated as the junction is illuminated with a 532 nm diode laser.  The $I_{ds}-V_{ds}$ curves at laser powers $0-10 \mu$W are shown in Fig. 3d.  The photocurrent, $I_{ph} = I_{ds,light} - I_{ds,dark}$, at $V_{ds}=-2$ V and a linear fit of $I_{ph}$ up to 8 $\mu$W are shown in the inset.  The slope from the fit gives a responsivity of 210 mA/W, which is comparable to commercial silicon photodetectors for green light.  Phototransistors from monolayer dichalcogenides can achieve even higher responsivities, though this improved responsivity seems to come at the expense of detector response time\cite{LopezSanchez:2013}.

In addition to photodetection, monolayer WSe$_2$ PN diodes are also capable of photovoltaic power generation.  With the device in the NP configuration, current is measured as a function of $V_{ds}$ for various laser powers from a 700 nm band-pass filtered supercontinuum white light source (Fig.~4a).  The short-circuit current, $I_{sc}$, which is the zero-bias current through the illuminated device, increases linearly with power up to at least 10 $\mu$W (Fig.~4a inset).  The power generated by the photovoltaic device, $P=I_{ds} \cdot V_{ds}$, is shown for laser powers $0-10 \mu$W in Fig.~4b.  The photovoltaic power generation also has a linear dependence on laser power (see Supplementary Information).

\begin{figure}[h]
\centering
\includegraphics[width=5.25in]{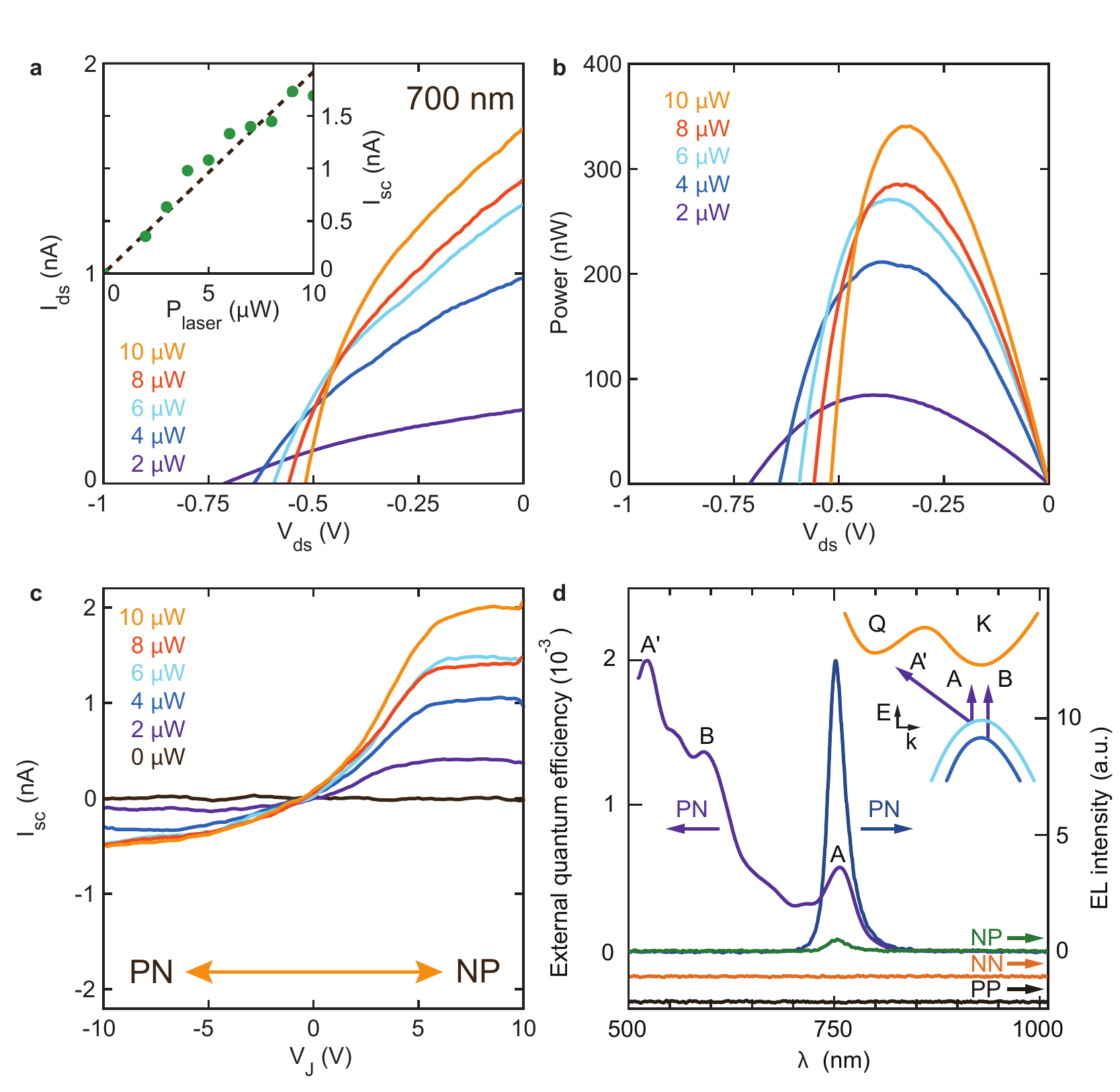}
\caption{
\textsf{\textbf{Photovoltaic response and light emission.  a,} $I_{ds}$ as a function of $V_{ds}$ in the NP configuration for laser powers $2-10 \mu$W (wavelength 700 nm).  Positive $I_{ds}$ and negative $V_{ds}$ in this regime reflects photovoltaic power generation.  Inset: Short circuit current, $I_{sc}$ (green dots), versus laser power with a linear fit (black dashed line).  \textbf{b,} Power, $P = I_{ds} \cdot V_{ds}$, produced by the device as a function of $V_{ds}$ for different incident laser powers, calculated from the data in \textbf{a}.  \textbf{c,} $I_{sc}$ as a function of $V_J$, for different laser powers.  $I_{sc}$ is nearly linear with power up to 10 $\mu$W in the NP configuration, though it saturates in the PN configuration beyond 6 $\mu$W.  As a function of gate voltage, the current saturates in both configurations for $V_J\sim \pm 5$ V.  \textbf{d,} Left axis:  External quantum efficiency as a function of wavelength at a constant laser power of 2 $\mu$W.  Peaks in the external quantum efficiency correspond to exciton transitions A, B, and A$^\prime$, as labeled.  Right axis:  Electroluminescence (EL) intensity at V$_{ds}=2$ V in the PN, NN, and PP configurations and at V$_{ds}=-2$ V in the NP configuration for a second monolayer WSe$_2$ device with one Au and one Pd contact.  NN and PP traces are offset vertically for clarity.  Inset: Diagram of the band structure around the K and Q points, with arrows indicating the lowest energy exciton transitions for monolayer WSe$_2$.
}}
\end{figure}

Varying $V_J$, the asymmetric gate voltage that defines the junction, at different laser powers, we observe a saturation in the short circuit current when $V_J \sim \pm 5$ V (Fig. 4c).  The current is higher for the NP configuration than for PN due to a difference in contact resistances between the two contacts.  We also note that the photocurrent due to the photovoltaic effect in these WSe$_2$ diodes is approximately an order of magnitude larger than photothermoelectric currents observed at the contacts of a monolayer MoS$_2$ field-effect transistor \cite{Buscema:2013}.

To obtain spectrally resolved photocurrent in the NP configuration, we measure $I_{sc}$ as a function of excitation wavelength.  To quantify the efficiency with which our device converts light into current at a particular wavelength, we extracted the external quantum efficiency, EQE $= (I_{sc}/P_{laser})(hc/e\lambda)$, as a function of wavelength, $\lambda$, at constant laser power, $P_{laser}$, where $h$, $c$, and $e$, are Planck's constant, speed of light, and electron charge, respectively.  We observe three peaks in the spectrally resolved photocurrent at 755, 591, and 522 nm (Fig. 4d), corresponding to energies of 1.64, 2.10, and 2.38 eV, respectively, with a maximum EQE of 0.2\% at 522 nm.  These energies match well with the values predicted\cite{Ding:2011} and experimentally observed via photoluminescence,\cite{Zhao:2013} differential reflectance,\cite{Zeng:2013} and optical absorption spectroscopy\cite{Huang:2013} for the A, B, and A$^\prime$ transitions of monolayer WSe$_2$, as depicted in the band diagram in the inset to Fig.~4d.

Finally, we measure the electroluminescence spectrum of a second monolayer WSe$_2$ PN diode (Fig.~4d).  To improve hole injection, this device was fabricated with Pd instead of Au in contact with the p-type region of the WSe$_2$ (see Supplementary Information).  With the device in the PN configuration, $V_{ds}=2$ V and $I_{ds}=100$ nA, the device behaves as a light-emitting diode (LED).  The emitted light has a spectral peak at 752 nm, which corresponds with the direct gap exciton transition seen in the photocurrent spectrum (Fig.~4d).  Light emission and a peak at $\sim 750$ nm can also be seen in the NP configuration ($V_{ds}=-2$ V and $I_{ds}=4$ nA), with a peak height smaller in proportion to the reduced $I_{ds}$.  No emission is seen in either the NN ($V_{ds}=2$ V and $I_{ds}=300$ nA) or PP ($V_{ds}=2$ V and $I_{ds}=500$ nA) configurations, confirming that the gate-defined PN junction generates the electroluminescence.

Based on the performance of the devices presented here, we anticipate a prominent role for diodes and optoelectronic devices based on monolayer dichalcogenide PN junctions.  Taking into account the three-atom thickness and low optical absorption of monolayer WSe$_2$ (Refs.~\citenum{Huang:2013,Bernardi:2013}) the responsivity and EQE reported here are quite substantial.  Additionally, the device geometry could be optimized to significantly enhance photoresponse.  In the split gate geometry studied here, the active region of the device is only a narrow line confined near the gap between the gates.  So only a fraction of the incident power from the laser spot contributes to photocurrent generation.  We expect that a vertical junction structure based on transfer-aligned exfoliated flakes\cite{Dean:2010} or large-area dichalcogenides grown by chemical vapor deposition\cite{Huang:2013} could increase responsivity and external quantum efficiency by more than an order of magnitude.  Further, improved contact to the dichalcogenide, particularly for holes, should dramatically improve device performance.

In conclusion, we have demonstrated electrically tunable PN diodes based solely on monolayer WSe$_2$.  These diodes strongly rectify current, in a direction selectable by the voltage settings of the two gates controlling the device.  Both the PN and NP configurations have diode ideality factors of $n=1.9$ and a rectification factor of 10$^5$.  When laser light is incident on the junction, these diodes produce a large photocurrent with a responsivity of $210$ mA/W at high bias.  At low bias, the diodes generate power via the photovoltaic effect, with a peak external quantum efficiency of 0.2\% at 522 nm.  The spectral response of the photocurrent from visible to near infrared wavelengths showed peaks corresponding to the three lowest excitonic transitions expected for monolayer WSe$_2$.  Finally, these devices can also function as light-emitting diodes with an electroluminescence peak at 752 nm.  These PN junction diodes demonstrate the potential of monolayer WSe$_2$, in addition to other direct gap semiconducting dichalcogenides, for novel electronic and optoelectronic applications.  As device quality improves, they also lay the foundation for more fundamental quantum transport experiments\cite{Li:2013}.

{\it Note added:} During the final preparation of this manuscript we became aware of similar work on PN diodes in monolayer WSe$_2$ (Ref.~\citenum{Pospischil:2013}).

\subsection*{Methods}
Device fabrication begins with exfoliation of bulk, natural WSe$_2$ (Nanosurf Inc.) down to few-layered sheets using the mechanical cleavage method pioneered for graphene\cite{Novoselov:2005}.  The thin flakes are deposited onto a glass/PDMS/MMA transfer slide as described in graphene-boron-nitride device fabrication\cite{Dean:2010,Zomer:2011}, and single molecular layers were identified by optical contrast\cite{Benameur:2011} and confirmed with AFM and either photocurrent spectroscopy or electroluminescence (see Supplementary Information).  These monolayers are then transferred onto a pair of split gates covered by 20 nm of HfO$_2$ (grown by atomic layer deposition).  The gates are separated by a 100 nm gap, patterned using e-beam lithography on a highly doped Si substrate covered in 285 nm of thermally grown SiO$_2$.  They are made from e-beam evaporated gold and are 20 nm thick.  The two back gates are capacitively coupled to the device through the HfO$_2$ dielectric, which has a dielectric constant, $\epsilon_r \approx 15$.
The WSe$_2$ is contacted by two gold electrodes, each approximately 1 $\mu$m wide and 25 nm thick with a 0.3 nm chromium sticking layer.

Electroluminescence was measured using a liquid nitrogen cooled CCD with an integration time of 60 seconds.  A background measured at $V_{ds}=0$ was subtracted from all four EL traces in Fig.~4d.

\subsection*{Acknowledgements}
We acknowledge experimental assistance from Kamphol Akkaravarawong, Trond Andersen, Nathaniel Gabor, Qian Lin, Qiong Ma, Wenjing Fang, and Javier Sanchez-Yamagishi and discussions with Patrick Brown.  This work has been primarily funded by ONR Young Investigator Award N00014-13-1-0610, and partly by the ONR GATE MURI and a Packard Fellowship.  This work made use of the MRSEC Shared Experimental Facilities supported by NSF under award No.~DMR-0819762 and of Harvard's CNS, supported by NSF under Grant ECS-0335765.

\subsection*{Author contributions}
B.W.H.B., H.O.H.C., and Y.Y. fabricated the samples, B.W.H.B., H.O.H.C., and Y.Y. performed the measurements, and all authors analyzed the data and co-wrote the paper.

\subsection*{Additional information}
Supplementary information is available in the online version of the paper.  Correspondence and requests for materials
should be addressed to P.J-H.~(\mbox{pjarillo@mit.edu}).


\begin{thebibliography}{35}
\expandafter\ifx\csname natexlab\endcsname\relax\def\natexlab#1{#1}\fi
\expandafter\ifx\csname bibnamefont\endcsname\relax
  \def\bibnamefont#1{#1}\fi
\expandafter\ifx\csname bibfnamefont\endcsname\relax
  \def\bibfnamefont#1{#1}\fi
\expandafter\ifx\csname citenamefont\endcsname\relax
  \def\citenamefont#1{#1}\fi
\expandafter\ifx\csname url\endcsname\relax
  \def\url#1{\texttt{#1}}\fi
\expandafter\ifx\csname urlprefix\endcsname\relax\def\urlprefix{URL }\fi
\providecommand{\bibinfo}[2]{#2}
\providecommand{\eprint}[2][]{\url{#2}}

\bibitem[{\citenamefont{Bertolazzi et~al.}(2011)\citenamefont{Bertolazzi,
  Brivio, and Kis}}]{Bertolazzi:2011}
\bibinfo{author}{\bibfnamefont{S.}~\bibnamefont{Bertolazzi}},
  \bibinfo{author}{\bibfnamefont{J.}~\bibnamefont{Brivio}}, \bibnamefont{and}
  \bibinfo{author}{\bibfnamefont{A.}~\bibnamefont{Kis}}, 
  \bibinfo{title}{Stretching and Breaking of Ultrathin MoS$_2$}. 
  \bibinfo{journal}{ACS
  Nano} \textbf{\bibinfo{volume}{5}}, \bibinfo{pages}{9703}
  (\bibinfo{year}{2011}).

\bibitem[{\citenamefont{Mak et~al.}(2010)\citenamefont{Mak, Lee, Hone, Shan,
  and Heinz}}]{Mak:2010}
\bibinfo{author}{\bibfnamefont{K.~F.} \bibnamefont{Mak}},
  \bibinfo{author}{\bibfnamefont{C.}~\bibnamefont{Lee}},
  \bibinfo{author}{\bibfnamefont{J.}~\bibnamefont{Hone}},
  \bibinfo{author}{\bibfnamefont{J.}~\bibnamefont{Shan}}, \bibnamefont{and}
  \bibinfo{author}{\bibfnamefont{T.~F.} \bibnamefont{Heinz}},
  \bibinfo{title}{Atomically Thin ${\mathrm{MoS}}_{2}$: A New Direct-Gap Semiconductor}. 
  \bibinfo{journal}{Phys. Rev. Lett.} \textbf{\bibinfo{volume}{105}},
  \bibinfo{pages}{136805} (\bibinfo{year}{2010}).

\bibitem[{\citenamefont{Chuang et~al.}(2013)\citenamefont{Chuang, Kapadia,
  Fang, Chang, Yen, Chueh, and Javey}}]{Chuang:2013}
\bibinfo{author}{\bibfnamefont{S.}~\bibnamefont{Chuang}},
  \bibinfo{author}{\bibfnamefont{R.}~\bibnamefont{Kapadia}},
  \bibinfo{author}{\bibfnamefont{H.}~\bibnamefont{Fang}},
  \bibinfo{author}{\bibfnamefont{T.~C.} \bibnamefont{Chang}},
  \bibinfo{author}{\bibfnamefont{W.-C.} \bibnamefont{Yen}},
  \bibinfo{author}{\bibfnamefont{Y.-L.} \bibnamefont{Chueh}}, \bibnamefont{and}
  \bibinfo{author}{\bibfnamefont{A.}~\bibnamefont{Javey}}, 
  \bibinfo{title}{Near-ideal electrical properties of InAs/WSe$_2$ van der Waals heterojunction diodes}. 
  \bibinfo{journal}{Applied Physics Letters} \textbf{\bibinfo{volume}{102}},
  \bibinfo{pages}{242101} (\bibinfo{year}{2013}).

\bibitem[{\citenamefont{Novoselov et~al.}(2005)\citenamefont{Novoselov, Jiang,
  Booth, Khotkevich, Morozov, and Geim}}]{Novoselov:2005}
\bibinfo{author}{\bibfnamefont{K.~S.} \bibnamefont{Novoselov}},
  \bibinfo{author}{\bibfnamefont{D.}~\bibnamefont{Jiang}},
  \bibinfo{author}{\bibfnamefont{T.}~\bibnamefont{Booth}},
  \bibinfo{author}{\bibfnamefont{V.~V.} \bibnamefont{Khotkevich}},
  \bibinfo{author}{\bibfnamefont{S.~M.} \bibnamefont{Morozov}},
  \bibnamefont{and} \bibinfo{author}{\bibfnamefont{A.~K.} \bibnamefont{Geim}}, 
  \bibinfo{title}{Two Dimensional Atomic Crystals}. 
  \bibinfo{journal}{Proc. Natl. Acad. Sci. U. S. A.}
  \textbf{\bibinfo{volume}{102}}, \bibinfo{pages}{10451}
  (\bibinfo{year}{2005}).

\bibitem[{\citenamefont{Zhao et~al.}(2013)\citenamefont{Zhao, Ghorannevis, Chu,
  Toh, Kloc, Tan, and Eda}}]{Zhao:2013}
\bibinfo{author}{\bibfnamefont{W.}~\bibnamefont{Zhao}},
  \bibinfo{author}{\bibfnamefont{Z.}~\bibnamefont{Ghorannevis}},
  \bibinfo{author}{\bibfnamefont{L.}~\bibnamefont{Chu}},
  \bibinfo{author}{\bibfnamefont{M.}~\bibnamefont{Toh}},
  \bibinfo{author}{\bibfnamefont{C.}~\bibnamefont{Kloc}},
  \bibinfo{author}{\bibfnamefont{P.-H.} \bibnamefont{Tan}}, \bibnamefont{and}
  \bibinfo{author}{\bibfnamefont{G.}~\bibnamefont{Eda}}, 
  \bibinfo{title}{Evolution of Electronic Structure in Atomically Thin Sheets of WS$_2$ and WSe$_2$}. 
  \bibinfo{journal}{ACS
  Nano} \textbf{\bibinfo{volume}{7}}, \bibinfo{pages}{791}
  (\bibinfo{year}{2013}).

\bibitem[{\citenamefont{Geim and Grigorieva}(2013)}]{Geim:2013}
\bibinfo{author}{\bibfnamefont{A.~K.} \bibnamefont{Geim}} \bibnamefont{and}
  \bibinfo{author}{\bibfnamefont{I.~V.} \bibnamefont{Grigorieva}}, 
  \bibinfo{title}{Van der Waals Heterostructures}. 
  \bibinfo{journal}{Nature} \textbf{\bibinfo{volume}{499}},
  \bibinfo{pages}{419} (\bibinfo{year}{2013}).

\bibitem[{\citenamefont{Wilson and Yoffe}(1969)}]{Wilson:1969}
\bibinfo{author}{\bibfnamefont{J.}~\bibnamefont{Wilson}} \bibnamefont{and}
  \bibinfo{author}{\bibfnamefont{A.}~\bibnamefont{Yoffe}}, 
  \bibinfo{title}{The transition metal dichalcogenides discussion and interpretation of the observed optical, electrical and structural properties}. 
  \bibinfo{journal}{Advances in Physics} \textbf{\bibinfo{volume}{18}},
  \bibinfo{pages}{193} (\bibinfo{year}{1969}).

\bibitem[{\citenamefont{Ding et~al.}(2011)\citenamefont{Ding, Wang, Ni, Shi,
  Shi, and Tang}}]{Ding:2011}
\bibinfo{author}{\bibfnamefont{Y.}~\bibnamefont{Ding}},
  \bibinfo{author}{\bibfnamefont{Y.}~\bibnamefont{Wang}},
  \bibinfo{author}{\bibfnamefont{J.}~\bibnamefont{Ni}},
  \bibinfo{author}{\bibfnamefont{L.}~\bibnamefont{Shi}},
  \bibinfo{author}{\bibfnamefont{S.}~\bibnamefont{Shi}}, \bibnamefont{and}
  \bibinfo{author}{\bibfnamefont{W.}~\bibnamefont{Tang}}, 
  \bibinfo{title}{First principles study of structural, vibrational and electronic properties of graphene-like MX$_2$}. 
  \bibinfo{journal}{Physica B: Condensed Matter}
  \textbf{\bibinfo{volume}{406}}, \bibinfo{pages}{2254 }
  (\bibinfo{year}{2011}).

\bibitem[{\citenamefont{Podzorov et~al.}(2004)\citenamefont{Podzorov,
  Gershenson, Kloc, Zeis, and Bucher}}]{Podzorov:2004}
\bibinfo{author}{\bibfnamefont{V.}~\bibnamefont{Podzorov}},
  \bibinfo{author}{\bibfnamefont{M.~E.} \bibnamefont{Gershenson}},
  \bibinfo{author}{\bibfnamefont{C.}~\bibnamefont{Kloc}},
  \bibinfo{author}{\bibfnamefont{R.}~\bibnamefont{Zeis}}, \bibnamefont{and}
  \bibinfo{author}{\bibfnamefont{E.}~\bibnamefont{Bucher}}, 
  \bibinfo{title}{High-mobility field-effect transistors based on transition metal dichalcogenides}. 
  \bibinfo{journal}{Applied Physics Letters} \textbf{\bibinfo{volume}{84}},
  \bibinfo{pages}{3301} (\bibinfo{year}{2004}).

\bibitem[{\citenamefont{Radisavljevic and Kis}(2013)}]{Radisavljevic:2013}
\bibinfo{author}{\bibfnamefont{B.}~\bibnamefont{Radisavljevic}}
  \bibnamefont{and} \bibinfo{author}{\bibfnamefont{A.}~\bibnamefont{Kis}}, 
  \bibinfo{title}{Mobility engineering and a metal–insulator transition in monolayer MoS$_2$}. 
  \bibinfo{journal}{Nature Materials} \textbf{\bibinfo{volume}{12}},
  \bibinfo{pages}{815} (\bibinfo{year}{2013}).

\bibitem[{\citenamefont{Baugher et~al.}(2013)\citenamefont{Baugher, Churchill,
  Yang, and Jarillo-Herrero}}]{Baugher:2013}
\bibinfo{author}{\bibfnamefont{B.~W.~H.} \bibnamefont{Baugher}},
  \bibinfo{author}{\bibfnamefont{H.~O.~H.} \bibnamefont{Churchill}},
  \bibinfo{author}{\bibfnamefont{Y.}~\bibnamefont{Yang}}, \bibnamefont{and}
  \bibinfo{author}{\bibfnamefont{P.}~\bibnamefont{Jarillo-Herrero}}, 
  \bibinfo{title}{Intrinsic Electronic Transport Properties of High-Quality Monolayer and Bilayer MoS$_2$}. 
  \bibinfo{journal}{Nano Letters} \textbf{\bibinfo{volume}{13}},
  \bibinfo{pages}{4212} (\bibinfo{year}{2013}).

\bibitem[{\citenamefont{Zhang et~al.}(2013)\citenamefont{Zhang, Ye, Yomogida,
  Takenobu, and Iwasa}}]{Zhang:2013}
\bibinfo{author}{\bibfnamefont{Y.~J.} \bibnamefont{Zhang}},
  \bibinfo{author}{\bibfnamefont{J.~T.} \bibnamefont{Ye}},
  \bibinfo{author}{\bibfnamefont{Y.}~\bibnamefont{Yomogida}},
  \bibinfo{author}{\bibfnamefont{T.}~\bibnamefont{Takenobu}}, \bibnamefont{and}
  \bibinfo{author}{\bibfnamefont{Y.}~\bibnamefont{Iwasa}}, 
  \bibinfo{title}{Formation of a Stable p–n Junction in a Liquid-Gated MoS$_2$ Ambipolar Transistor}. 
  \bibinfo{journal}{Nano Letters} \textbf{\bibinfo{volume}{13}},
  \bibinfo{pages}{3023} (\bibinfo{year}{2013}).

\bibitem[{\citenamefont{Wang et~al.}(2012)\citenamefont{Wang, Yu, Lee, Shi,
  Hsu, Chin, Li, Dubey, Kong, and Palacios}}]{Wang:2012}
\bibinfo{author}{\bibfnamefont{H.}~\bibnamefont{Wang}},
  \bibinfo{author}{\bibfnamefont{L.}~\bibnamefont{Yu}},
  \bibinfo{author}{\bibfnamefont{Y.-H.} \bibnamefont{Lee}},
  \bibinfo{author}{\bibfnamefont{Y.}~\bibnamefont{Shi}},
  \bibinfo{author}{\bibfnamefont{A.}~\bibnamefont{Hsu}},
  \bibinfo{author}{\bibfnamefont{M.~L.} \bibnamefont{Chin}},
  \bibinfo{author}{\bibfnamefont{L.-J.} \bibnamefont{Li}},
  \bibinfo{author}{\bibfnamefont{M.}~\bibnamefont{Dubey}},
  \bibinfo{author}{\bibfnamefont{J.}~\bibnamefont{Kong}}, \bibnamefont{and}
  \bibinfo{author}{\bibfnamefont{T.}~\bibnamefont{Palacios}}, 
  \bibinfo{title}{Integrated Circuits Based on Bilayer MoS$_2$ Transistors}. 
  \bibinfo{journal}{Nano Letters} \textbf{\bibinfo{volume}{12}},
  \bibinfo{pages}{4674} (\bibinfo{year}{2012}).

\bibitem[{\citenamefont{Yin et~al.}(2012)\citenamefont{Yin, Li, Li, Jiang, Shi,
  Sun, Lu, Zhang, Chen, and Zhang}}]{Yin:2012}
\bibinfo{author}{\bibfnamefont{Z.}~\bibnamefont{Yin}},
  \bibinfo{author}{\bibfnamefont{H.}~\bibnamefont{Li}},
  \bibinfo{author}{\bibfnamefont{H.}~\bibnamefont{Li}},
  \bibinfo{author}{\bibfnamefont{L.}~\bibnamefont{Jiang}},
  \bibinfo{author}{\bibfnamefont{Y.}~\bibnamefont{Shi}},
  \bibinfo{author}{\bibfnamefont{Y.}~\bibnamefont{Sun}},
  \bibinfo{author}{\bibfnamefont{G.}~\bibnamefont{Lu}},
  \bibinfo{author}{\bibfnamefont{Q.}~\bibnamefont{Zhang}},
  \bibinfo{author}{\bibfnamefont{X.}~\bibnamefont{Chen}}, \bibnamefont{and}
  \bibinfo{author}{\bibfnamefont{H.}~\bibnamefont{Zhang}}, 
  \bibinfo{title}{Single-Layer MoS$_2$ Phototransistors}. 
  \bibinfo{journal}{ACS Nano} \textbf{\bibinfo{volume}{6}}, \bibinfo{pages}{74}
  (\bibinfo{year}{2012}).

\bibitem[{\citenamefont{Lopez-Sanchez et~al.}(2013)\citenamefont{Lopez-Sanchez,
  Lembke, Kayci, Radenovic, and Kis}}]{LopezSanchez:2013}
\bibinfo{author}{\bibfnamefont{O.}~\bibnamefont{Lopez-Sanchez}},
  \bibinfo{author}{\bibfnamefont{D.}~\bibnamefont{Lembke}},
  \bibinfo{author}{\bibfnamefont{M.}~\bibnamefont{Kayci}},
  \bibinfo{author}{\bibfnamefont{A.}~\bibnamefont{Radenovic}},
  \bibnamefont{and} \bibinfo{author}{\bibfnamefont{A.}~\bibnamefont{Kis}}, 
  \bibinfo{title}{Ultrasensitive photodetectors based on monolayer MoS$_2$}. 
  \bibinfo{journal}{Nature Nanotechnology} \textbf{\bibinfo{volume}{8}},
  \bibinfo{pages}{497} (\bibinfo{year}{2013}).

\bibitem[{\citenamefont{Splendiani et~al.}(2010)\citenamefont{Splendiani, Sun,
  Zhang, Li, Kim, Chim, Galli, and Wang}}]{Splendiani:2010}
\bibinfo{author}{\bibfnamefont{A.}~\bibnamefont{Splendiani}},
  \bibinfo{author}{\bibfnamefont{L.}~\bibnamefont{Sun}},
  \bibinfo{author}{\bibfnamefont{Y.}~\bibnamefont{Zhang}},
  \bibinfo{author}{\bibfnamefont{T.}~\bibnamefont{Li}},
  \bibinfo{author}{\bibfnamefont{J.}~\bibnamefont{Kim}},
  \bibinfo{author}{\bibfnamefont{C.-Y.} \bibnamefont{Chim}},
  \bibinfo{author}{\bibfnamefont{G.}~\bibnamefont{Galli}}, \bibnamefont{and}
  \bibinfo{author}{\bibfnamefont{F.}~\bibnamefont{Wang}}, 
  \bibinfo{title}{Emerging Photoluminescence in Monolayer MoS$_2$}. 
  \bibinfo{journal}{Nano Letters} \textbf{\bibinfo{volume}{10}},
  \bibinfo{pages}{1271} (\bibinfo{year}{2010}).

\bibitem[{\citenamefont{Sundaram et~al.}(2013)\citenamefont{Sundaram, Engel,
  Lombardo, Krupke, Ferrari, Avouris, and Steiner}}]{Sundaram:2012}
\bibinfo{author}{\bibfnamefont{R.~S.} \bibnamefont{Sundaram}},
  \bibinfo{author}{\bibfnamefont{M.}~\bibnamefont{Engel}},
  \bibinfo{author}{\bibfnamefont{A.}~\bibnamefont{Lombardo}},
  \bibinfo{author}{\bibfnamefont{R.}~\bibnamefont{Krupke}},
  \bibinfo{author}{\bibfnamefont{A.~C.} \bibnamefont{Ferrari}},
  \bibinfo{author}{\bibfnamefont{P.}~\bibnamefont{Avouris}}, \bibnamefont{and}
  \bibinfo{author}{\bibfnamefont{M.}~\bibnamefont{Steiner}}, 
  \bibinfo{title}{Electroluminescence in Single Layer MoS$_2$}. 
  \bibinfo{journal}{Nano Letters} \textbf{\bibinfo{volume}{13}},
  \bibinfo{pages}{1416} (\bibinfo{year}{2013}).

\bibitem[{\citenamefont{Ye et~al.}(2013)\citenamefont{Ye, Ye, Gharghi, Zhu,
  Zhao, Yin, and Zhang}}]{Ye:2013}
\bibinfo{author}{\bibfnamefont{Y.}~\bibnamefont{Ye}},
  \bibinfo{author}{\bibfnamefont{Z.}~\bibnamefont{Ye}},
  \bibinfo{author}{\bibfnamefont{M.}~\bibnamefont{Gharghi}},
  \bibinfo{author}{\bibfnamefont{H.}~\bibnamefont{Zhu}},
  \bibinfo{author}{\bibfnamefont{M.}~\bibnamefont{Zhao}},
  \bibinfo{author}{\bibfnamefont{X.}~\bibnamefont{Yin}}, \bibnamefont{and}
  \bibinfo{author}{\bibfnamefont{X.}~\bibnamefont{Zhang}}, 
  \bibinfo{title}{Exciton-related electroluminescence from monolayer MoS$_2$}. 
  \bibinfo{journal}{arXiv:1305.4235}  (\bibinfo{year}{2013}).

\bibitem[{\citenamefont{Zeng et~al.}(2013)\citenamefont{Zeng, Liu, Dai, Yan,
  Zhu, He, Xie, Xu, Chen, Yao et~al.}}]{Zeng:2013}
\bibinfo{author}{\bibfnamefont{H.}~\bibnamefont{Zeng}},
  \bibinfo{author}{\bibfnamefont{G.-B.} \bibnamefont{Liu}},
  \bibinfo{author}{\bibfnamefont{J.}~\bibnamefont{Dai}},
  \bibinfo{author}{\bibfnamefont{Y.}~\bibnamefont{Yan}},
  \bibinfo{author}{\bibfnamefont{B.}~\bibnamefont{Zhu}},
  \bibinfo{author}{\bibfnamefont{R.}~\bibnamefont{He}},
  \bibinfo{author}{\bibfnamefont{L.}~\bibnamefont{Xie}},
  \bibinfo{author}{\bibfnamefont{S.}~\bibnamefont{Xu}},
  \bibinfo{author}{\bibfnamefont{X.}~\bibnamefont{Chen}},
  \bibinfo{author}{\bibfnamefont{W.}~\bibnamefont{Yao}}, \bibnamefont{et~al.}, 
  \bibinfo{title}{Optical signature of symmetry variations and spin-valley coupling in atomically thin tungsten dichalcogenides}. 
  \bibinfo{journal}{Sci. Rep.} \textbf{\bibinfo{volume}{3}}
  (\bibinfo{year}{2013}).

\bibitem[{\citenamefont{Mak et~al.}(2012)\citenamefont{Mak, He, Shan, and
  Heinz}}]{Mak:2012}
\bibinfo{author}{\bibfnamefont{K.~F.} \bibnamefont{Mak}},
  \bibinfo{author}{\bibfnamefont{K.}~\bibnamefont{He}},
  \bibinfo{author}{\bibfnamefont{J.}~\bibnamefont{Shan}}, \bibnamefont{and}
  \bibinfo{author}{\bibfnamefont{T.~F.} \bibnamefont{Heinz}}, 
  \bibinfo{title}{Control of valley polarization in monolayer MoS$_2$ by optical helicity}. 
  \bibinfo{journal}{Nature Nanotechnology} \textbf{\bibinfo{volume}{7}},
  \bibinfo{pages}{494} (\bibinfo{year}{2012}).

\bibitem[{\citenamefont{Sallen et~al.}(2012)\citenamefont{Sallen, Bouet, Marie,
  Wang, Zhu, Han, Lu, Tan, Amand, Liu et~al.}}]{Sallen:2012}
\bibinfo{author}{\bibfnamefont{G.}~\bibnamefont{Sallen}},
  \bibinfo{author}{\bibfnamefont{L.}~\bibnamefont{Bouet}},
  \bibinfo{author}{\bibfnamefont{X.}~\bibnamefont{Marie}},
  \bibinfo{author}{\bibfnamefont{G.}~\bibnamefont{Wang}},
  \bibinfo{author}{\bibfnamefont{C.~R.} \bibnamefont{Zhu}},
  \bibinfo{author}{\bibfnamefont{W.~P.} \bibnamefont{Han}},
  \bibinfo{author}{\bibfnamefont{Y.}~\bibnamefont{Lu}},
  \bibinfo{author}{\bibfnamefont{P.~H.} \bibnamefont{Tan}},
  \bibinfo{author}{\bibfnamefont{T.}~\bibnamefont{Amand}},
  \bibinfo{author}{\bibfnamefont{B.~L.} \bibnamefont{Liu}},
  \bibnamefont{et~al.}, 
  \bibinfo{title}{Robust optical emission polarization in MoS$_{2}$ monolayers through selective valley excitation}. 
  \bibinfo{journal}{Phys. Rev. B}
  \textbf{\bibinfo{volume}{86}}, \bibinfo{pages}{081301}
  (\bibinfo{year}{2012}).

\bibitem[{\citenamefont{Jones et~al.}(2013)\citenamefont{Jones, Yu, Ghimire,
  Wu, Aivazian, Ross, Zhao, Yan, Mandrus, and Xiao}}]{Jones:2013}
\bibinfo{author}{\bibfnamefont{A.~M.} \bibnamefont{Jones}},
  \bibinfo{author}{\bibfnamefont{H.}~\bibnamefont{Yu}},
  \bibinfo{author}{\bibfnamefont{N.}~\bibnamefont{Ghimire}},
  \bibinfo{author}{\bibfnamefont{S.}~\bibnamefont{Wu}},
  \bibinfo{author}{\bibfnamefont{G.}~\bibnamefont{Aivazian}},
  \bibinfo{author}{\bibfnamefont{J.~S.} \bibnamefont{Ross}},
  \bibinfo{author}{\bibfnamefont{B.}~\bibnamefont{Zhao}},
  \bibinfo{author}{\bibfnamefont{J.}~\bibnamefont{Yan}},
  \bibinfo{author}{\bibfnamefont{D.}~\bibnamefont{Mandrus}}, \bibnamefont{and}
  \bibinfo{author}{\bibfnamefont{D.}~\bibnamefont{Xiao}}, 
  \bibinfo{title}{Optical Generation of Excitonic Valley Coherence in Monolayer WSe$_2$}. 
  \bibinfo{journal}{Nature Nanotechnology} \textbf{\bibinfo{volume}{8}},
  \bibinfo{pages}{634} (\bibinfo{year}{2013}).

\bibitem[{\citenamefont{Spah et~al.}(1983)\citenamefont{Spah, Elrod,
  Luxsteiner, Bucher, and Wagner}}]{Spah:1983}
\bibinfo{author}{\bibfnamefont{R.}~\bibnamefont{Spah}},
  \bibinfo{author}{\bibfnamefont{U.}~\bibnamefont{Elrod}},
  \bibinfo{author}{\bibfnamefont{M.}~\bibnamefont{Luxsteiner}},
  \bibinfo{author}{\bibfnamefont{E.}~\bibnamefont{Bucher}}, \bibnamefont{and}
  \bibinfo{author}{\bibfnamefont{S.}~\bibnamefont{Wagner}}, 
  \bibinfo{title}{Pn Junctions in Tungsten Diselenide}. 
  \bibinfo{journal}{Applied Physics Letters} \textbf{\bibinfo{volume}{43}},
  \bibinfo{pages}{79} (\bibinfo{year}{1983}).

\bibitem[{\citenamefont{Benameur et~al.}(2011)\citenamefont{Benameur,
  Radisavljevic, Héron, Sahoo, Berger, and Kis}}]{Benameur:2011}
\bibinfo{author}{\bibfnamefont{M.~M.} \bibnamefont{Benameur}},
  \bibinfo{author}{\bibfnamefont{B.}~\bibnamefont{Radisavljevic}},
  \bibinfo{author}{\bibfnamefont{J.~S.} \bibnamefont{Héron}},
  \bibinfo{author}{\bibfnamefont{S.}~\bibnamefont{Sahoo}},
  \bibinfo{author}{\bibfnamefont{H.}~\bibnamefont{Berger}}, \bibnamefont{and}
  \bibinfo{author}{\bibfnamefont{A.}~\bibnamefont{Kis}}, 
  \bibinfo{title}{Visibility of dichalcogenide nanolayers}. 
  \bibinfo{journal}{Nanotechnology} \textbf{\bibinfo{volume}{22}},
  \bibinfo{pages}{125706} (\bibinfo{year}{2011}).

\bibitem[{\citenamefont{Dean et~al.}(2010)\citenamefont{Dean, Young, Meric,
  Lee, Wang, Sorgenfrei, Watanabe, Taniguchi, Kim, Shepard et~al.}}]{Dean:2010}
\bibinfo{author}{\bibfnamefont{C.~R.} \bibnamefont{Dean}},
  \bibinfo{author}{\bibfnamefont{A.~F.} \bibnamefont{Young}},
  \bibinfo{author}{\bibfnamefont{I.}~\bibnamefont{Meric}},
  \bibinfo{author}{\bibfnamefont{C.}~\bibnamefont{Lee}},
  \bibinfo{author}{\bibfnamefont{L.}~\bibnamefont{Wang}},
  \bibinfo{author}{\bibfnamefont{S.}~\bibnamefont{Sorgenfrei}},
  \bibinfo{author}{\bibfnamefont{K.}~\bibnamefont{Watanabe}},
  \bibinfo{author}{\bibfnamefont{T.}~\bibnamefont{Taniguchi}},
  \bibinfo{author}{\bibfnamefont{P.}~\bibnamefont{Kim}},
  \bibinfo{author}{\bibfnamefont{K.~L.} \bibnamefont{Shepard}},
  \bibnamefont{et~al.}, 
  \bibinfo{title}{Boron nitride substrates for high-quality graphene electronics}. 
  \bibinfo{journal}{Nat Nano}
  \textbf{\bibinfo{volume}{5}}, \bibinfo{pages}{722} (\bibinfo{year}{2010}).

\bibitem[{\citenamefont{Zomer et~al.}(2011)\citenamefont{Zomer, Dash, Tombros,
  and van Wees}}]{Zomer:2011}
\bibinfo{author}{\bibfnamefont{P.~J.} \bibnamefont{Zomer}},
  \bibinfo{author}{\bibfnamefont{S.~P.} \bibnamefont{Dash}},
  \bibinfo{author}{\bibfnamefont{N.}~\bibnamefont{Tombros}}, \bibnamefont{and}
  \bibinfo{author}{\bibfnamefont{B.~J.} \bibnamefont{van Wees}}, 
  \bibinfo{title}{A transfer technique for high mobility graphene devices on commercially available hexagonal boron nitride}. 
  \bibinfo{journal}{Applied Physics Letters} \textbf{\bibinfo{volume}{99}},
  \bibinfo{pages}{232104} (\bibinfo{year}{2011}).

\bibitem[{\citenamefont{Radisavljevic et~al.}(2011)\citenamefont{Radisavljevic,
  Radenovic, Brivio, Giacometti, and Kis}}]{Radisavljevic:2011}
\bibinfo{author}{\bibfnamefont{B.}~\bibnamefont{Radisavljevic}},
  \bibinfo{author}{\bibfnamefont{A.}~\bibnamefont{Radenovic}},
  \bibinfo{author}{\bibfnamefont{J.}~\bibnamefont{Brivio}},
  \bibinfo{author}{\bibfnamefont{V.}~\bibnamefont{Giacometti}},
  \bibnamefont{and} \bibinfo{author}{\bibfnamefont{A.}~\bibnamefont{Kis}}, 
  \bibinfo{title}{Single-layer MoS2 transistors}. 
  \bibinfo{journal}{Nature Nanotechnology} \textbf{\bibinfo{volume}{6}},
  \bibinfo{pages}{147} (\bibinfo{year}{2011}).

\bibitem[{\citenamefont{Banwell and Jayakumar}(2000)}]{Banwell:2000}
\bibinfo{author}{\bibfnamefont{T.}~\bibnamefont{Banwell}} \bibnamefont{and}
  \bibinfo{author}{\bibfnamefont{A.}~\bibnamefont{Jayakumar}}, 
  \bibinfo{title}{Exact analytical solution for current flow through diode with series resistance}. 
  \bibinfo{journal}{Electronics Letters} \textbf{\bibinfo{volume}{36}},
  \bibinfo{pages}{291} (\bibinfo{year}{2000}).

\bibitem[{\citenamefont{Sah et~al.}(1957)\citenamefont{Sah, Noyce, and
  Shockley}}]{Sah:1957}
\bibinfo{author}{\bibfnamefont{C.-T.} \bibnamefont{Sah}},
  \bibinfo{author}{\bibfnamefont{R.~N.} \bibnamefont{Noyce}}, \bibnamefont{and}
  \bibinfo{author}{\bibfnamefont{W.}~\bibnamefont{Shockley}}, 
  \bibinfo{title}{Carrier generation and recombination in p-n junctions and p-n junction characteristics}. 
  \bibinfo{journal}{Proceedings of the IRE} \textbf{\bibinfo{volume}{45}},
  \bibinfo{pages}{1228} (\bibinfo{year}{1957}).

\bibitem[{\citenamefont{Lambert}(1758)}]{Lambert:1758}
\bibinfo{author}{\bibfnamefont{J.}~\bibnamefont{Lambert}}, 
  \bibinfo{title}{Observationes variae in Mathes in Puram}. 
  \bibinfo{journal}{Acta Helvetica,
  Physico-mathematico-anatomico-botanico-medica} \textbf{\bibinfo{volume}{3}},
  \bibinfo{pages}{128} (\bibinfo{year}{1758}).

\bibitem[{\citenamefont{Buscema et~al.}(2013)\citenamefont{Buscema, Barkelid,
  Zwiller, and van~der Zant}}]{Buscema:2013}
\bibinfo{author}{\bibfnamefont{M.}~\bibnamefont{Buscema}},
  \bibinfo{author}{\bibfnamefont{M.}~\bibnamefont{Barkelid}},
  \bibinfo{author}{\bibfnamefont{V.}~\bibnamefont{Zwiller}}, \bibnamefont{and}
  \bibinfo{author}{\bibfnamefont{H.}~\bibnamefont{van~der Zant}}, 
  \bibinfo{title}{Large and tunable photo-thermoelectric effect in single-layer MoS$_2$}. 
  \bibinfo{journal}{Nano Letters} \textbf{\bibinfo{volume}{13}},
  \bibinfo{pages}{358} (\bibinfo{year}{2013}).

\bibitem[{\citenamefont{Huang et~al.}(2013)\citenamefont{Huang, Pu, Chuu, Hsu,
  Chiu, Juang, Chang, Chang, Iwasa, and Chou}}]{Huang:2013}
\bibinfo{author}{\bibfnamefont{J.-K.} \bibnamefont{Huang}},
  \bibinfo{author}{\bibfnamefont{J.}~\bibnamefont{Pu}},
  \bibinfo{author}{\bibfnamefont{C.-P.} \bibnamefont{Chuu}},
  \bibinfo{author}{\bibfnamefont{C.-L.} \bibnamefont{Hsu}},
  \bibinfo{author}{\bibfnamefont{M.-H.} \bibnamefont{Chiu}},
  \bibinfo{author}{\bibfnamefont{Z.-Y.} \bibnamefont{Juang}},
  \bibinfo{author}{\bibfnamefont{Y.-H.} \bibnamefont{Chang}},
  \bibinfo{author}{\bibfnamefont{W.-H.} \bibnamefont{Chang}},
  \bibinfo{author}{\bibfnamefont{Y.}~\bibnamefont{Iwasa}}, \bibnamefont{and}
  \bibinfo{author}{\bibfnamefont{M.-Y.} \bibnamefont{Chou}}, 
  \bibinfo{title}{Large-Area and Highly Crystalline WSe$_2$ Monolayers: from Synthesis to Device Applications}. 
  \bibinfo{journal}{arXiv:1304.7365}  (\bibinfo{year}{2013}).

\bibitem[{\citenamefont{Bernardi et~al.}(2013)\citenamefont{Bernardi, Palummo,
  and Grossman}}]{Bernardi:2013}
\bibinfo{author}{\bibfnamefont{M.}~\bibnamefont{Bernardi}},
  \bibinfo{author}{\bibfnamefont{M.}~\bibnamefont{Palummo}}, \bibnamefont{and}
  \bibinfo{author}{\bibfnamefont{J.~C.} \bibnamefont{Grossman}}, 
  \bibinfo{title}{Exraordinary Sunlight Absorption and One Nanometer Thick Photovoltaics Using Two-dimensional Monolayer Materials}. 
  \bibinfo{journal}{Nano Letters} \textbf{\bibinfo{volume}{13}},
  \bibinfo{pages}{3664} (\bibinfo{year}{2013}).

\bibitem[{\citenamefont{Li et~al.}(2013)\citenamefont{Li, Zhang, and
  Niu}}]{Li:2013}
\bibinfo{author}{\bibfnamefont{X.}~\bibnamefont{Li}},
  \bibinfo{author}{\bibfnamefont{F.}~\bibnamefont{Zhang}}, \bibnamefont{and}
  \bibinfo{author}{\bibfnamefont{Q.}~\bibnamefont{Niu}}, 
  \bibinfo{title}{Unconventional Quantum Hall Effect and Tunable Spin Hall Effect in Dirac Materials: Application to an Isolated MoS$_{2}$ Trilayer}. 
  \bibinfo{journal}{Physical Review Letters} \textbf{\bibinfo{volume}{110}},
  \bibinfo{pages}{066803} (\bibinfo{year}{2013}).

\bibitem[{\citenamefont{Pospischil et~al.}(2013)\citenamefont{Pospischil,
  Furchi, and Mueller}}]{Pospischil:2013}
\bibinfo{author}{\bibfnamefont{A.}~\bibnamefont{Pospischil}},
  \bibinfo{author}{\bibfnamefont{M.~M.} \bibnamefont{Furchi}},
  \bibnamefont{and} \bibinfo{author}{\bibfnamefont{T.}~\bibnamefont{Mueller}}, 
  \bibinfo{title}{Solar energy conversion and light emission in an atomic monolayer p-n diode}. 
  \bibinfo{journal}{arXiv:1309.7492v1}  (\bibinfo{year}{2013}).

\end{thebibliography}
\end{document}